\documentclass{aa}  

\usepackage{hyperref}

\usepackage{graphicx}
\usepackage{txfonts}
\begin{document}

   \title{3D Alfv\'en Wave Behaviour Around \\ Proper and Improper Magnetic Null Points}

  \author{J.~O. Thurgood
          \and
          J.~A. McLaughlin
          }

   \institute{Department of Mathematics \& Information Sciences, Northumbria University, Newcastle Upon Tyne, NE1 8ST, UK \\
              \email{\href{mailto:jonathan.thurgood@northumbria.ac.uk}{\nolinkurl{jonathan.thurgood@northumbria.ac.uk}  }}             }

   \date{Received MONTH DATE, YEAR; accepted MONTH DATE, YEAR}

 
  \abstract
   {MHD waves and magnetic null points are both prevalent in many astrophysical plasmas, including the solar atmosphere. Interaction between waves and null points has been implicated as a possible mechanism for localised heating events.}
   { Here  we investigate the transient behaviour of the Alfv\'en wave about fully 3D proper and improper 3D  magnetic null points. Previously, the behaviour of fast magnetoacoustic waves at null points in 3D, cold MHD was  considered by Thurgood \& McLaughlin 
(\href{http://dx.doi.org/10.1051/0004-6361/201219850}{Astronomy \& Astrophysics, 2012, 545, A9}).}
   {We introduce an Alfv\'en wave into the vicinity of both proper and improper null points by numerically solving the ideal, $\beta=0$ MHD equations using the LARE3D code. A magnetic fieldline and flux-based coordinate system permits the isolation of resulting wave-modes and the analysis of their interaction.}
   {We find that the Alfv\'en wave propagates throughout the region and accumulates near the fan-plane, causing current build up. For different values of null point  eccentricity, the qualitative behaviour changes only by the imposition of anisotropic pulse dilation, due to the differing rates at which fieldlines diverge from the spine. For all eccentricities, we find that the fast and Alfv\'en waves are linearly decoupled. During the driving phase, an independently propagating fast wave is nonlinearly generated due to the ponderomotive force. Subsequently, no further excitation of fast waves occurs.
   }
   {We find that the key aspects of the theory of Alfv\'en waves about  2D null points extends intuitively to the fully 3D case; i.e. the wave propagates along fieldlines and thus accumulates at predictable parts of the topology. We also highlight that unlike in the 2D case, in 3D Alfv\'en-wave pulses \emph{are always toroidal} and thus any aspects of 2D Alfv\'en-wave-null models that are pulse-geometry specific must be reconsidered in 3D.}

   \keywords{magnetohydrodynamics (MHD) -- waves--  Sun: corona -- Sun: oscillations --Sun: magnetic topology  -- magnetic fields}

   \maketitle
%

\section{Introduction}
Over the past few decades, high-resolution and high-cadence instruments aboard satellites such as SOHO, TRACE, Hinode and SDO have revealed that MHD waves and oscillations are abundant throughout the coronal plasma and are present in most, if not all, of its structures (see reviews by, eg. De Moortel \citeyear{ineke2005}; Nakariakov \& Verwichte \citeyear{NK2005}; Ruderman \& Erd\'elyi \citeyear{Ruderman2009}; Goosens et al. \citeyear{goosens2011}; and references therein). 
Such observations have lead to a rapid development of MHD wave theory, perhaps most notably regarding the dynamics of waves in coronal loops for the purposes of \emph{coronal seismology} (e.g., De Moortel \& Nakariakov \citeyear{ineke2012}). However prevalent, these curvilinear structures form only part of a wider set of topological features. Since the behaviour of MHD waves strongly depends on the structure of the background magnetic field, the behaviour of waves can vary greatly in different topologies. To fully understand the oscillatory corona, it is necessary to develop wave theory for the non-curvilinear and inhomogeneous magnetic topologies topologies also.    In this paper we are concerned with MHD wave dynamics in one such topology and investigate the behaviour of such waves in  the vicinity of  \emph{magnetic null points.}

Magnetic null points are singular locations in the field where magnetic induction is zero, and are a natural consequence of multiple sources of magnetic flux. The existence of magnetic null points in the coronal magnetic field is implied by field extrapolations (such as, e.g., Brown \& Priest \citeyear{BrownPriest01}; 
Beveridge et al. \citeyear{Beveridge2002}) and have been estimated to number $1.0$-$4.0\times10^{4}$ points in the corona (Close et al. \citeyear{closeparnellpriest04}; R\'egnier et al. \citeyear{regnierparnellhaynes08}; Longcope \& Parnell \citeyear{loncopeparnell09}). Outside of the corona, the Cluster mission has detected null points in the Earth's magnetotail (eg. Xiao et al. \citeyear{Xiao2007}), and modelling suggests clusters of nulls exist in the global magnetosphere (eg. Dorelli et al. \citeyear{Dorelli2007}).
Null points have also been identified as playing key roles in many processes of the solar atmosphere, such as; in magnetic reconnection (see, e.g., Pontin \citeyear{Pontin2012}), oscillatory reconnection (McLaughlin et al. \citeyear{JamesOR2009}; \citeyear{JamesOR2012}; Threlfall \citeyear{Threl2012}) and in CMEs (in the \emph{magnetic breakout model}, e.g. Antiochos \citeyear{antichos98}; Antiochos et al. \citeyear{antichos99}). Given the ubiquity of both MHD waves and the prevalence of null points, wave-null interactions are an inevitable fundamental aspect of the dynamic solar atmosphere.

The behaviour of MHD waves in the vicinity of magnetic null points has been studied extensively using 2D models. Here we seek to evaluate the extent to which the key results regarding transient behaviour of waves extend to the 3D case, thus we only discuss the 2D case briefly to highlight  fundamental results. For a review of 2D null point investigations, see McLaughlin et al. (\citeyear{JAMESNULLREVIEW}).

 A series of papers by McLaughlin \& Hood (\citeyear{MH2004}; \citeyear{MH2005}; \citeyear{MH2006a}) considered the transient behaviour of Alfv\'en and fast magnetoacoustic waves about null points in various ideal, $\beta=0$ 2D scenarios and found that the differing modes of oscillation always have distinct features; the fast wave \emph{always refracts} along the Alfv\'en-speed profile, accumulating at the null point; whereas the Alfv\'en wave is confined to fieldlines, and accumulates along separatricies which it \emph{cannot cross}. Over time, regardless of the initial wave configuration, all of the wave's energy will accumulate at these regions of topology, with ever steepening gradients. These regions are thus identified as locations for preferential heating by (passing) MHD waves, via ohmic heating. Furthermore, the authors found that  waves of different modes do not interact (in their linear solution). 
A later nonlinear solution of the same scenario (Thurgood \& McLaughlin \citeyear{Me2013_NL2DAWAVE}) considered the possibility of nonlinear mode coupling due to the inhomogeneous geometry of the 2D null. When an Alfv\'en wave propagates in regions of non-uniform Alfv\'en-speed profile, magnetoacoustic waves can be excited via the nonlinear magnetic pressure gradients (viz. \emph{ponderomotive force} see, e.g., Nakariakov et al. \citeyear{NK97}; Verwichte et al. \citeyear{Erwin99}; Botha et al. \citeyear{gert2000}; Tsiklauri et al. \citeyear{tsiklauri2001}; Thurgood \& McLaughlin \citeyear{Me2013_PMF}).  Thurgood \& McLaughlin (\citeyear{Me2013_NL2DAWAVE}) found that at 2D nulls such mode excitation only occurred during driving at boundaries, and subsequently   no further excitation occurred despite the inhomogeneous field. 
%
Overall, they determined that effect did not significantly impact upon the dynamics of the main Alfv\'en wave. As such, the original conclusions of McLaughlin \& Hood(\citeyear{MH2005}; \citeyear{MH2005}; \citeyear{MH2006a}) hold in the shock-free  nonlinear case and at 2D null points in the cold-plasma limit.

However, as singularities, null points are intrinsically 3D, thus the 2D configurations discussed in fact capture the physics of a \emph{null line} of infinite extent. The 2D studies therefore serve to give an initial grounding in the physics of realistic null points, and for a more complete understanding we must consider the 3D case. 

Until recently, papers that addressed the topic of MHD wave behaviour about 3D null points primarily focused on aspects of current accumulation (in an attempt to identify regions where reconnection is likely to occur) rather than the transient features of MHD wave propagation.
Notably, studies by Galsgaard et al. (\citeyear{klaus03}), Pontin \& Galsgaard (\citeyear{pontingalsgaard07}), Pontin et al. (\citeyear{pontinetal07}) and Galsgaard \& Pontin (\citeyear{galsgaard11b}; \citeyear{galsgaard11a}) considered various driving motions on boundaries which resulted, depending on the particular nature of driving, in current accumulation at the null point, the spine or the fan plane (characteristic regions of 3D null topology, see \S \ref{section:2.2}). 
Thurgood \& McLaughlin (\citeyear{Me2012a}, their section 1) suggested that these papers represented a tantalising suggestion that, like their 2D counter parts, at 3D nulls MHD waves behaved in a discrete, decoupled manner and that they accumulated at predictable regions of the magnetic topology. They considered the nature of the fast magnetoacoustic wave in the 3D regime for null points of varying eccentricity (see \S \ref{section:2.2}). 
The study confirmed that in all cases that fast magnetoacoustic waves eventually accumulate at the null point due to refraction along the Alfv\'en speed profile (albeit at different rates, due to the effect of the field eccentricity). They concluded that, \emph{3D null points are likely locations of localised heating events due to passing fast waves { being trapped in the vicinity}.} Additionally, they found no evidence of geometric or nonlinear coupling to the Alfv\'en mode, although the propagating wave did sustain a nonlinear disturbance longitudinally to the background field (which was later confirmed as due to the action of the ponderomotive force in Thurgood \& McLaughlin \citeyear{Me2013_PMF}). 
Finally, we note that the observed behaviour of the fast wave strongly corresponded to that predicted by the WKB-method as considered in McLaughlin et al. (\citeyear{james083dwkb}). 

McLaughlin et al. (\citeyear{james083dwkb}) also derived a WKB approximation for the behaviour of the Alfv\'en wave at fully 3D null points. Their findings strongly suggest that, in 3D, the Alfv\'en wave is confined to fieldlines, propagates at the equilibrium Alfv\'en speed and  accumulates along the spine and fan of the null point (an intuitive extension of the 2D results). However, their implementation of the WKB method was unable to address the possible geometric or nonlinear interaction between modes.

Thus, as it stands, the question of how MHD waves behave in the neighbourhood of fully 3D magnetic null points is only partially answered. In this paper, we consider the complementary scenario to that of Thurgood \& McLaughlin (\citeyear{Me2012a}) and investigate the dynamics of the Alfv\'en wave about 3D null points of varying eccentricity 
{(as opposed to the behaviour of the fast wave at different 3D null points)}.
Thus, we consider this work to be a companion paper to Thurgood \& McLaughlin (\citeyear{Me2012a}), hereafter refereed to as Paper 1.

We proceed follows: in \S \ref{section:2} we outline the methods used to model the scenario (governing equations \S \ref{section:2.1}, null point topology \S \ref{section:2.2}, the method for isolating wave modes \S \ref{section:2.3}, and numerical solution \S \ref{section:2.4}); in \S \ref{section:3} we detail the results at the proper (\S \ref{section:3.1}) and improper (\S \ref{section:3.2}) null points. Finally,  in \S \ref{section:4} we present our conclusions and then also we discuss the combined implications of the results of this paper and Paper 1 in \S \ref{section:4.1}.

\section{Mathematical Model}\label{section:2}

We model the behaviour of the Alfv\'en wave at 3D null points in an analogous way to the process previously used to model the behaviour of the fast wave at 3D null points. Here, we summarise our modelling methods. For further details, see sections 2-2.5 of Paper 1.

\begin{figure*}
\centering
\includegraphics[width=17cm]{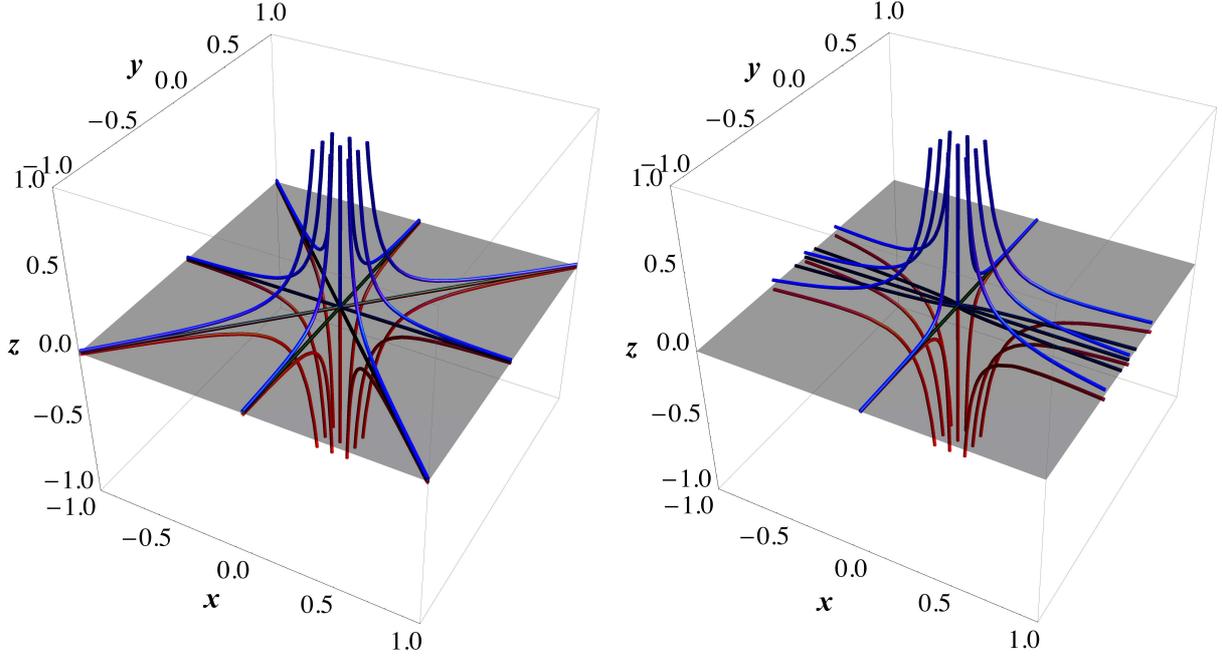}
\caption{Left: Indicative fieldlines for the $\epsilon=1$, azimuthally-symmetric proper null point. Right: the same fieldlines for the $\epsilon=0.5$ improper null point (note the loss of rotational symmetry).  Blue fieldlines originate from $z>0$, red from $z<0$ and the black fieldlines originate at the null and lie in the fan plane, shaded grey. }
\label{fig:fields}
\end{figure*}

\subsection{Governing Equations} \label{section:2.1}

The three-dimensional, nonlinear, ideal, adiabatic $\beta=0$ MHD equations are considered. The $\beta=0$ approximation is used to prohibit acoustic effects (chiefly to prohibit the introduction of the  slow mode) in order to restrict our focus to the behaviour of Alfv\'en wave and its interplay with the fast magnetoacoustic mode. The governing equations are as follows:
\begin{eqnarray}
\rho\left[\frac{\partial\mathbf{v}}{\partial t}+(\mathbf{v}\cdot\mathbf{\nabla})\mathbf{v}\right]&=& \left(\frac{\mathbf{\nabla}\times\mathbf{B}}{\mu} \right)\times\mathbf{B}\quad,\nonumber \\
\frac{\partial\mathbf{B}}{\partial t}&=&\mathbf{\nabla}\times(\mathbf{v}\times\mathbf{B})\quad,\nonumber\\
\frac{\partial\rho}{\partial t}&=& -\mathbf{\nabla}\cdot(\rho\mathbf{v})  \quad,\nonumber\\
\frac{\partial p}{\partial t} &=& - \mathbf{v}\cdot\nabla p-\gamma p \mathbf{\nabla}\cdot\mathbf{v}\quad.
\label{MHDeqns}
\end{eqnarray}
Here, standard MHD notation applies: $\mathbf{v}$ is plasma velocity, $p$ is thermal pressure, $\rho$ is density, $\mathbf{B}$ is the magnetic field/induction,  $\gamma=5/3$ is the adiabatic index, and $\mu$ is the magnetic permeability.

We consider an equilibrium state of $\rho=\rho_0$, $p=p_0$ (where $\rho_0$ and $p_0$ are constants), ${\mathbf{v}}=\mathbf{0}$ and  equilibrium magnetic field ${\bf{B}}=\mathbf{B}_0$. Finite, small perturbations of amplitude $\alpha \ll 1$ are considered in the form
$\rho=\rho_0 + \alpha\rho_1(\mathbf{r},t) $, $p=p_0 + \alpha p_1(\mathbf{r},t)$, $\mathbf{v}=\mathbf{0}+\alpha\mathbf{v}(\mathbf{r},t)$ and $\mathbf{B}=\mathbf{B}_0 +\alpha\mathbf{b}(\mathbf{r},t)$
and a subsequent nondimensionalisation using the substitution $\mathbf{v}=\overline{v}\mathbf{v}^*$,$\nabla={\nabla^*}/{L}$, $\mathbf{B}_{0}=B_{0}\mathbf{B}_{0}^{*}$, $\mathbf{b}=B_{0}\mathbf{b}^{*}$, $\mathbf{t}=\overline{t}\mathbf{t}^{*}$, $p_1=p_{0}p_{1}^{*}$ and $\rho_1=\rho_{0}\rho_{1}^{*}$ is performed, with the additional choices $\overline{v}={L} / {\overline{t}}$ and $\overline{v}={B_{0}} / {\sqrt{\mu\rho_{0}}}$.  The resulting nondimensionalised, governing equations of the perturbed system are:
\begin{eqnarray}
\frac{\partial \mathbf{v}}{\partial t} &=& \left(\nabla\times\mathbf{b}\right)\times\mathbf{B}_0 + \mathbf{N}_{1} \nonumber\\
\frac{\partial\mathbf{b}}{\partial t} &=& \nabla\times\left( \mathbf{v} \times\mathbf{B}_{0}\right) +  \mathbf{N}_{2} \nonumber\\
\frac{\partial \rho_1}{\partial t} &=& -\nabla\cdot \mathbf{v} + {N}_{3}\nonumber\\
\frac{\partial p_1}{\partial t} &=& -\gamma \nabla \cdot \mathbf{v} + {N}_{4} \nonumber\\
{\bf{N}}_1 &=& \left(\nabla\times\mathbf{b}\right)\times\mathbf{b} -\rho_{1}\frac{\partial \mathbf{v}}{\partial t}  - \left(1+\rho_{0}\right)\left(\mathbf{v}\cdot\nabla\right)\mathbf{v} \nonumber \\
{\bf{N}}_{2} &=& \nabla \times \left(\mathbf{v}\times\mathbf{b}\right)\nonumber \\
{{N}}_3 &=& -\nabla \cdot \left(\rho_{1} \mathbf{v} \right)  \nonumber \\
{{N}}_4 &=& \mathbf{v} \cdot \left(\nabla p_1 \right) -\gamma p_{1} \left(\nabla\cdot\mathbf{v}\right)\label{equation_MHD}
\end{eqnarray}
where terms ${\bf{N}}_i$ are the nonlinear components and the star indices have been dropped, henceforth all equations are presented in a nondimensional form.

\subsection{Equilibrium magnetic field: potential 3D null point }\label{section:2.2}

Here we consider the behaviour of Alfv\'en waves in the vicinity of single potential null points of the form:
\begin{eqnarray}
\mathbf{B}_{0}=\left[x,\epsilon y, -\left(\epsilon+1\right)z\right]  \label{eqn:3D_null_point}
\end{eqnarray}
where  the eccentricity parameter $\epsilon$ controls the direction in which fieldlines  predominantly align, and the null point itself is located at the origin. Two key features of the 3D null point are the {\emph{spine}} line and the {\emph{fan}} plane (see Figure 1 and Priest \& Titov \citeyear{priesttitov}).
The spine is an isolated fieldline along the $z$-axis that approaches, or leaves, the null point. In this paper, without loss of generality, we restrict our attention to {\emph{positive}} null points ($\epsilon\ge0$), and as such the spine represents fieldlines approaching the null from above and below the $z=0$ plane. The $z=0$ plane, known as the fan, consists of radial fieldlines confined to the plane that point away from the null point (for $\epsilon\ge0$).

Altering the eccentricity parameter $\epsilon$ changes the field topology as follows:
\def\labelitemi{$\bullet$}
\begin{itemize}
\item For $\epsilon=1$, the magnetic null point has azimuthal symmetry about its spine, with no preferred direction for fieldlines, and is known as a {\emph{proper}} null. 
\item Null points which deviate from this cylindrical symmetry are known as {\emph{improper}} nulls. For $0\leq\epsilon\leq1$ the fieldlines curve to run primarily parallel to the $x$-axis, and for $\epsilon\geq1$ curve to run parallel to the $y$-axis.
\item For $\epsilon=0$, we recover the simple 2D null point in the $xz-$plane, with a null line running through $x=z=0$.
\end{itemize}
For more comprehensive information on the classification of different types of 3D null, see Parnell et al. (\citeyear{parnell96}) .

\subsection{Isolating MHD modes}\label{section:2.3}

In Paper 1, we developed a magnetic-flux based coordinate system where each direction corresponds to a distinct MHD wave mode. 
Here, the system is used in two ways. Firstly, it is used to construct driving conditions that introduce linearly pure Alfv\'en waves ({i.e. the alternative case to that considered in Paper 1, where fast waves are introduced}, see $\S \ref{section:2.4}$). Secondly, it is used to track interaction between different modes of oscillation during the simulation.

 For null points of the form ($\ref{eqn:3D_null_point}$) the coordinate system is of the following form: perturbations to fluid-variables in the $\hat{\mathbf{A}}$-direction are associated with Alfv\'en waves, in $\hat{\mathbf{C}}$ correspond to fast waves, and in $\hat{\mathbf{B}}_{0}$ correspond to longitudinal disturbances (the slow wave is absent in the $\beta=0$ limit); where:
\begin{equation}
\begin{array}{l}
\displaystyle {\mathbf{B}}_{0}=\left[x,\epsilon y,-(\epsilon+1)z\right]\\
\displaystyle \mathbf{A}=\left[zy,-\epsilon xz,(1-\epsilon)xy\right]\\
\displaystyle \mathbf{C}=\left[C_x,C_y,C_z\right] \\
\displaystyle C_x = x \left[ \left(\epsilon^2-\epsilon\right)y^{2}+\left(\epsilon^2+\epsilon\right)z^{2}\right] \\
\displaystyle C_y = y  \left[ \left(1-\epsilon \right)x^{2}+\left(\epsilon+1\right)z^{2}\right] \\
\displaystyle C_z = \epsilon z \left(x^{2}+y^{2}\right)
\end{array} 
\label{eq:gencoordsyst}
\end{equation}
with unit normals $\mathbf{\hat{B}}=\mathbf{B}/|\mathbf{B}|$, $\mathbf{\hat{A}}=\mathbf{A}/|\mathbf{A}|$ and $\mathbf{\hat{C}}=\mathbf{C}/|\mathbf{C}|$. Note that on the line of the spine ($x=y=0$) and the fan plane ($z=0$), $\mathbf{B} = \nabla \times \mathbf{A}$ no longer holds, rendering the system locally invalid and as such, {\emph{the spine or fan  cannot be used to drive pure modes}} under this coordinate system. This relates to a degeneracy between the fast and Alfv\'en wave in these specific regions. For a further explination of the coordinate system, which 	is applicable to cases other than the 3D null, see sections $2.3$ and $2.3.1$ of Paper 1. Note that there exists a purely typographical error (i.e., the results are not effected) for $C_x$ in Paper 1. The correct form of $C_x$ is given above.

\begin{figure*}
\centering
\includegraphics[width=17cm]{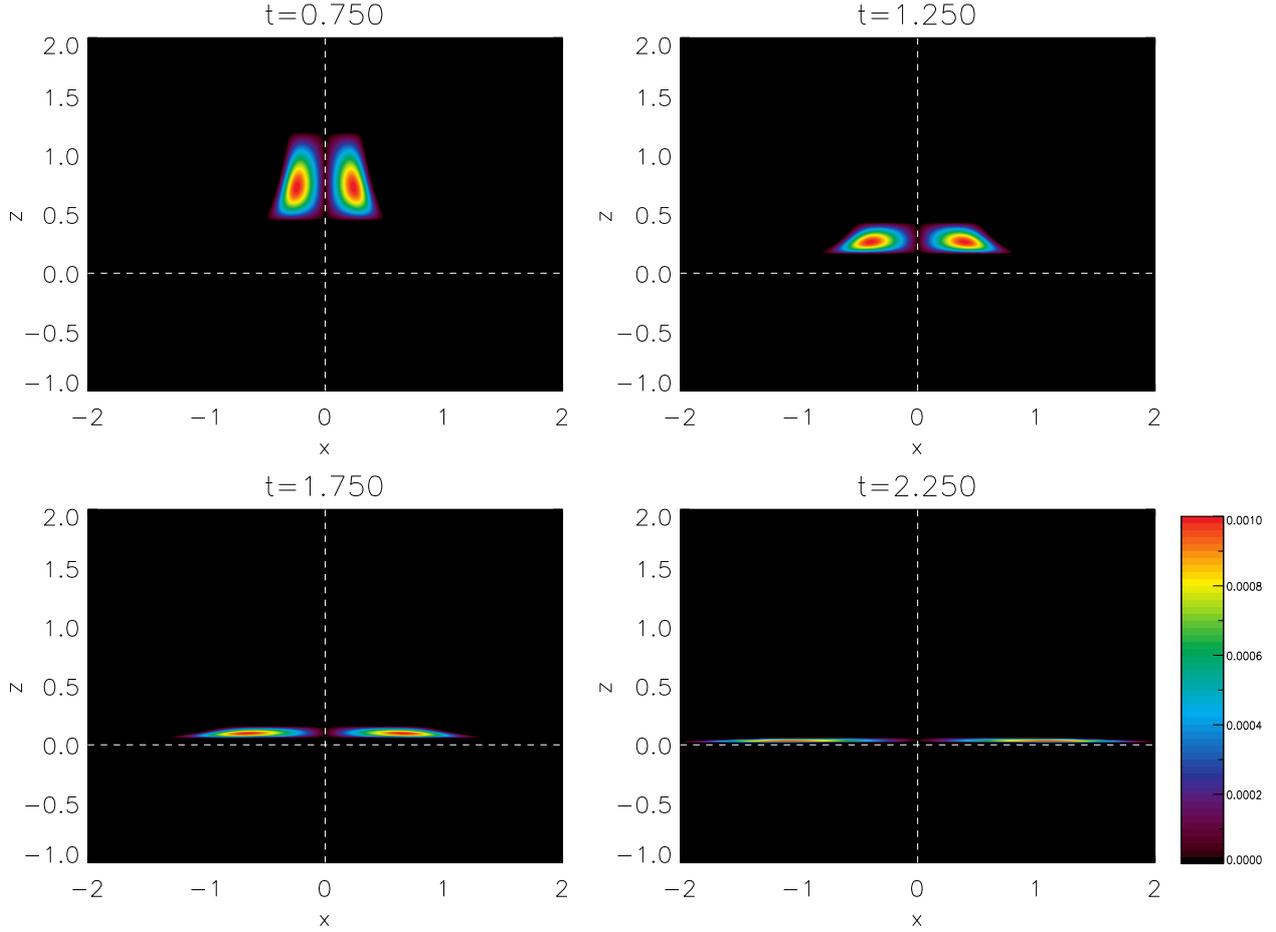}
\caption{The evolution of the Alfv\'en wave pulse at the proper null point, observed in $|{v_{A}}|$, in the $xz$-plane with $y=0$. Due to the azimuthal symmetry, the corresponding figures for other planes which intersect the spine-line $x=y=0$ are identical.  The dotted white lines indicate the position of the spine line and fan plane.}
\label{va_proper}
\end{figure*}

\begin{figure*}
\centering
\includegraphics[width=17cm]{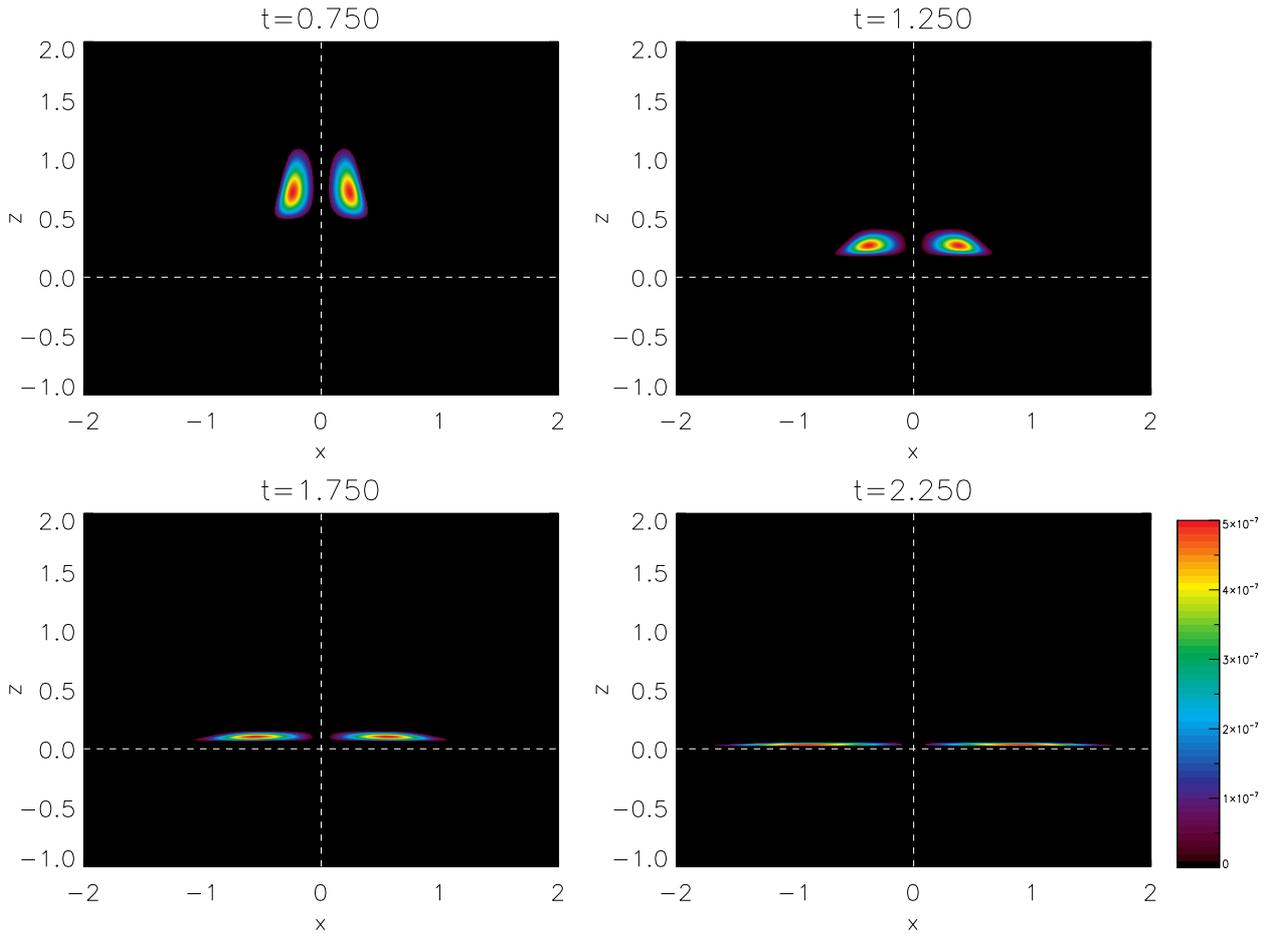}
\caption{The evolution of $|{v_{B}}|$ at the proper null, which shows a \emph{longitudinal daughter disturbance}, {in the $xz$-plane with $y=0$}. It is nonlinear and everywhere cospatial to the Alfv\'en wave. It arises due to the action of the Alfv\'en wave's ponderomotive force, which carries the disturbance along its path as it propagates. The daughter disturbance does not impact upon the medium as it passes through.}
\label{vB_proper}
\end{figure*}

\begin{figure*}
\centering
\includegraphics[width=17cm]{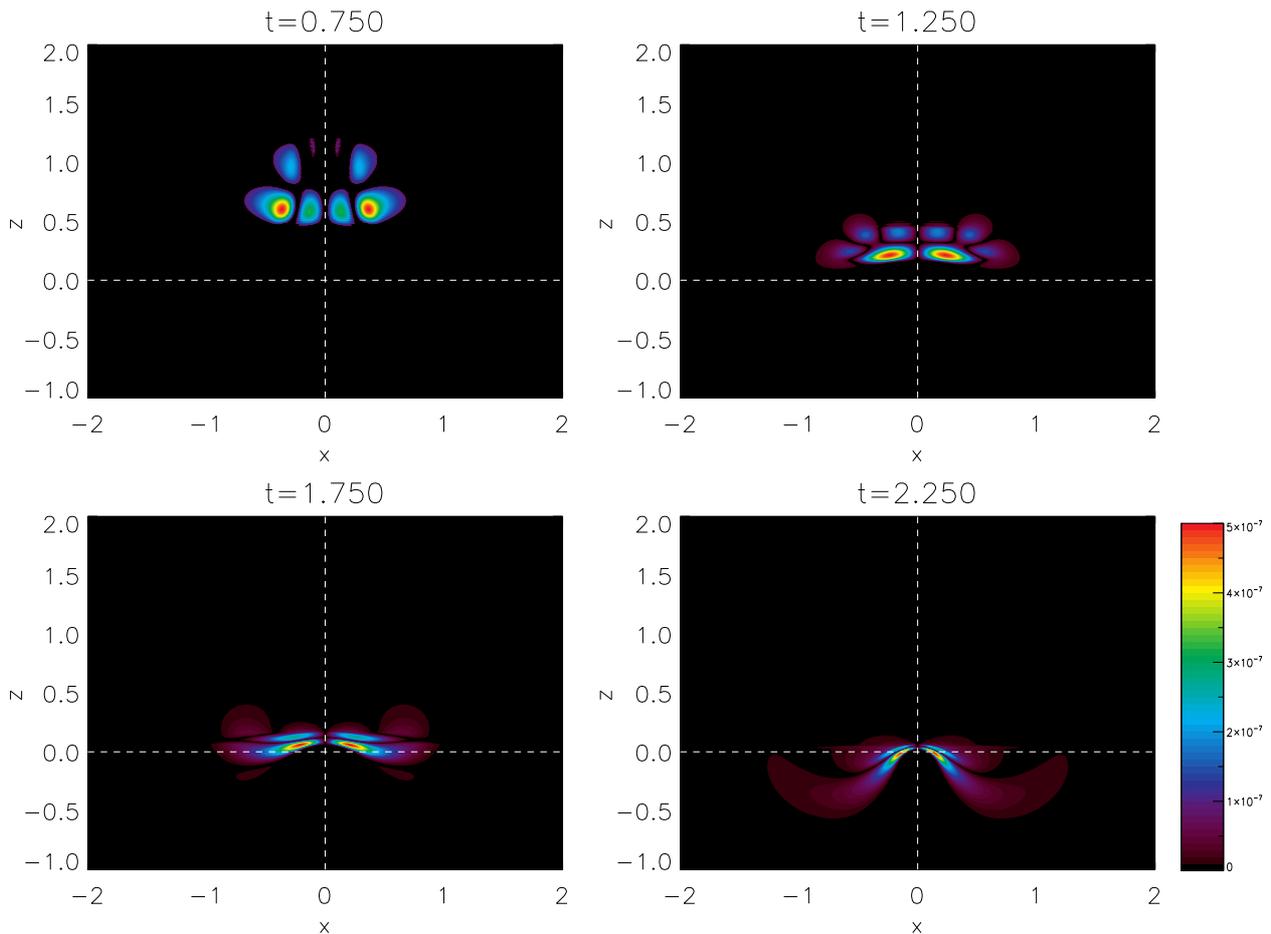}
\caption{The evolution of $|{v_{C}}|$ at the proper null, in which we see an independently propagating wave, which generated during the driving phase and is of nonlinear magnitude. It refracts along the Alfv\'en-speed profile, crosses the fan plane and accumulates at the null. }
\label{vc_proper}
\end{figure*}

\begin{figure*}
\centering
\includegraphics[width=17cm]{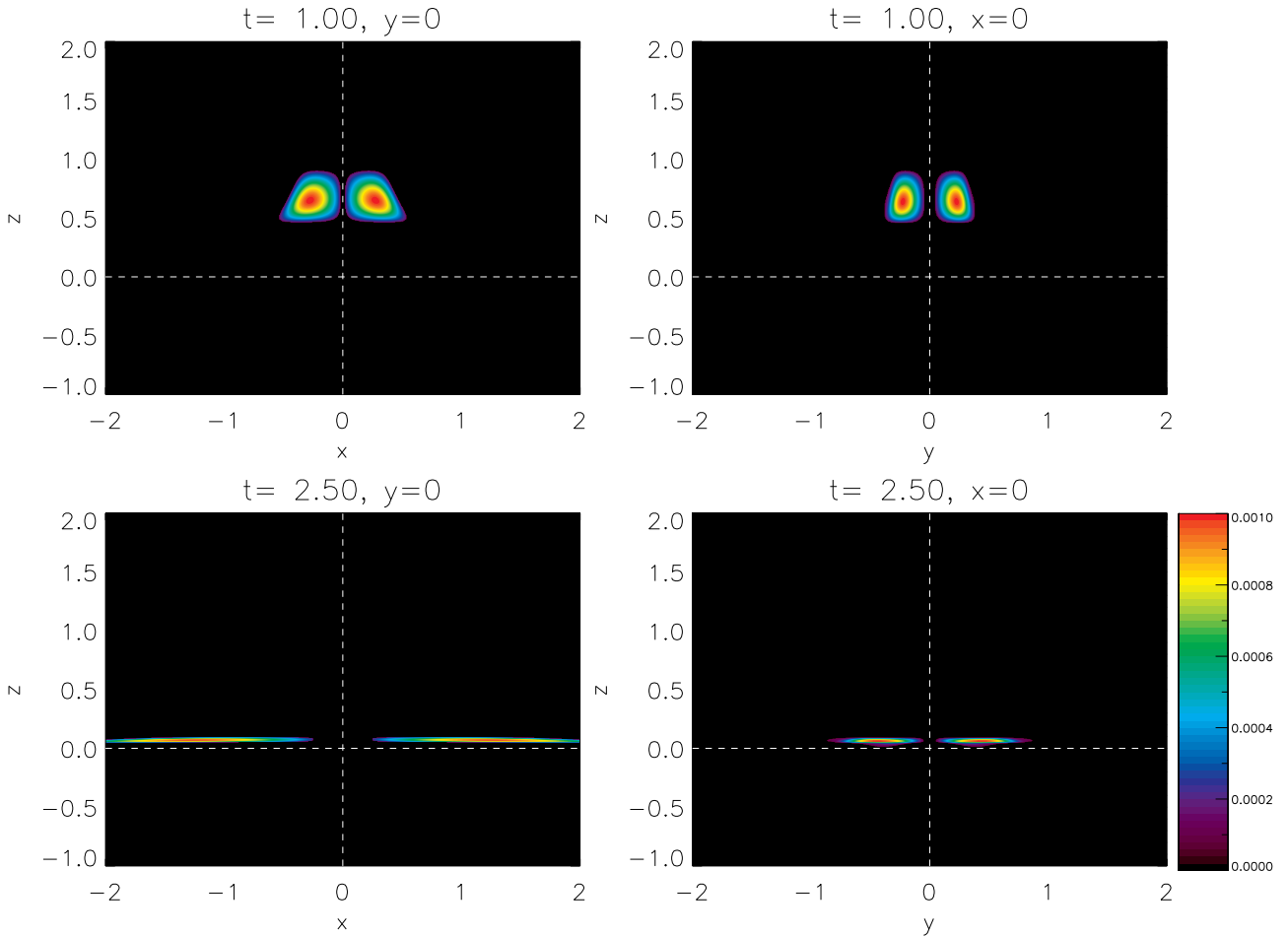}
\caption{The Alfv\'en wave, here shown in $|{v_{A}}|$, for the improper null $\epsilon=0.5$. As azimuthal symmetry is broken, we show two planes $y=0$ and $x=0$ (left and right columns respectively). {For this null, fieldlines predominantly align to run parallel to the  $x$-axis (see Figure \ref{fig:fields}); note that that the pulse spreads at different rates in the different planes.}}
\label{va_improper}
\end{figure*}

\begin{figure*}
\centering
\includegraphics[width=17cm]{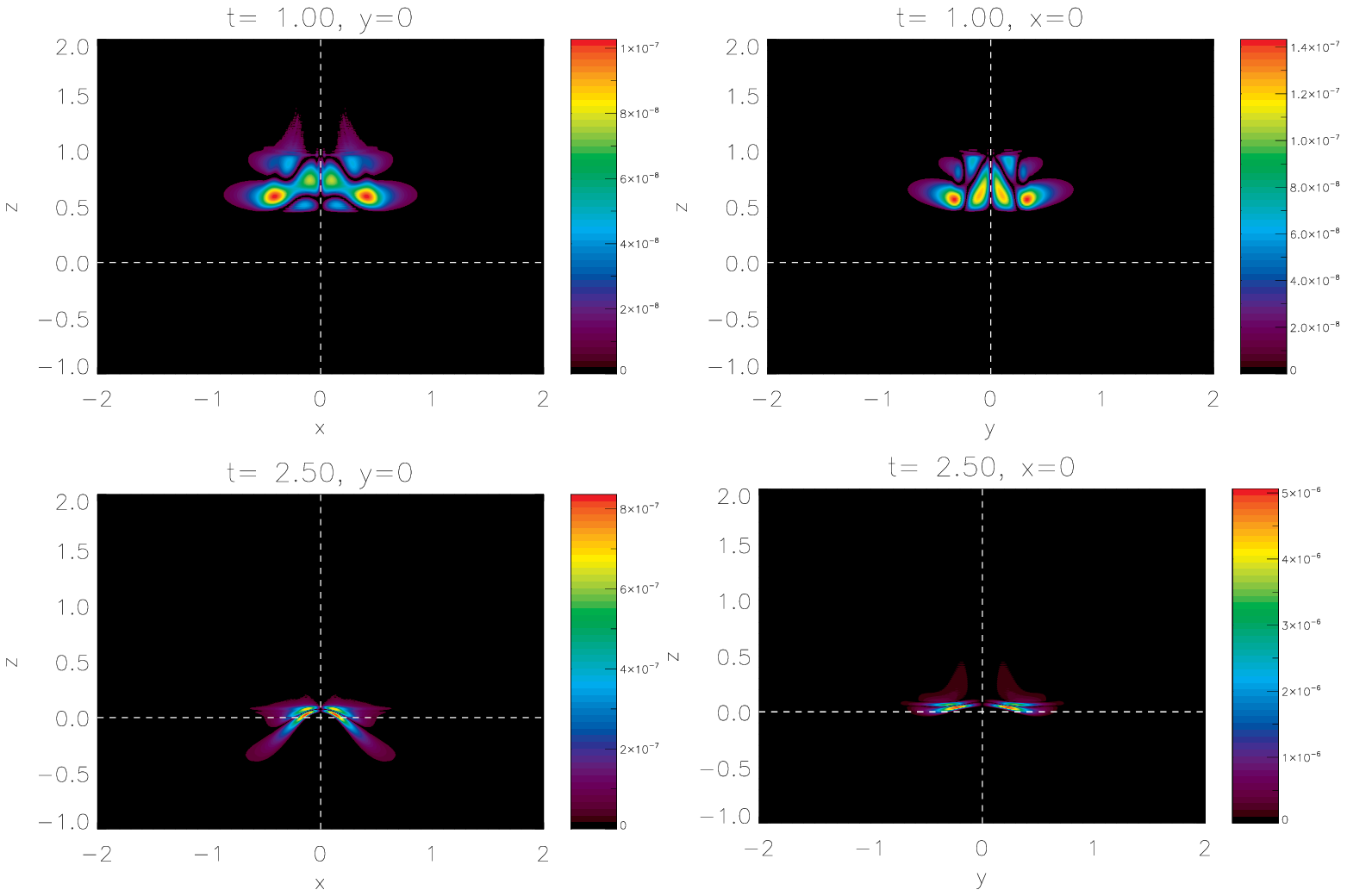}
\caption{The evolution of $|{v_{C}}|$ at the improper null $\epsilon=0.5$. An independently propagating fast magnetoacoustic wave is generated during the driving phase. Again, as azimuthal symmetry is lost we show two planes to demonstrate the different rates of refraction.}
\label{vc_improper}
\end{figure*}

\subsection{Numerical Solution}\label{section:2.4}

The fully nonlinear MHD equations (\ref{MHDeqns}) are solved using the {\emph{LARE3D}} numerical code (Arber et al. \citeyear{LAREpaper})  with magnetic equilibria corresponding to both proper and improper  3D null points ($\epsilon=1$ and $\epsilon=0.5$, see \S \ref{section:3.1} and \S \ref{section:3.2} respectively). In each scenario we introduce an Alfv\'en wave at the upper $z$-boundary, by driving the following: 
\begin{equation}
\mathbf{v}\cdot\hat{\mathbf{A}}=v_{A} = \alpha F(t)G(r) \;\;, \quad  
\mathbf{v}\cdot{\hat{\mathbf{B}}_0}=v_{B}=0  \;\;, \quad 
\mathbf{v}\cdot\hat{\mathbf{C}}=v_{C} =0
\label{pulse}
\end{equation}
\begin{equation*}
F(t)=\sin(2\pi t) \;\;, \quad G(r)=\frac{20}{9}\frac{r}{r_0}\left[1-\left(\frac{r}{r_0}\right)^4\right]^2 \;\;, \quad r=\sqrt{x^2 +y^2}
\end{equation*}
for $0\le t \le 0.5$, $0\le r \le r_0$, with driving amplitude $\alpha=0.001$ (we consider the weakly-nonlinear case). Note that $v_A$, $v_B$, and $v_C$ are velocity components in the directions of the coordinate system discussed in $\S \ref{section:2.3}$, and that $v_A$ is \emph{not} the Alfv\'en speed (which we denote $c_A$). The factor $\frac{20}{9}$ normalises function $G(r)$ such that the maxima/minima in the range is $\pm 1$. We choose $r_0=0.25$. The other boundary conditions are set as zero-gradient conditions. The simulations  utilise a uniform numerical grid with  domain $-3 \le x \le 3$, $-3 \le y \le 3$, $-1 \le z \le 2$ with $960 \times 960 \times 720$ grid points. The results presented focus into the region $-2 \le x \le 2$, $-2 \le y \le 2$, $-1 \le z \le 2$, i.e. a subset of the full numerical domain. The experiment ends just prior to the first instance of waves reaching the side boundaries (to ensure boundary reflection is not an issue).

\section{Numerical Results}\label{section:3}

\subsection{Proper Null, $\epsilon=1$}\label{section:3.1}

Let us first consider the proper, radial null point ($\epsilon=1$) where an Alfv\'en wave is introduced along the top boundary ($z=2$) by driving  $v_A$ according to equation \ref{pulse}.
The resultant propagation of the Alfv\'en wave is shown in Figure \ref{va_proper}. Figure \ref{va_proper} follows the Alfv\'en wave in the $xz$-plane with $y=0$, which is manifest in the velocity perturbation  in the invariant direction ${\hat{\mathbf{A}}}$ (NB: the corresponding field-perturbation $\mathbf{b}\cdot\hat{\mathbf{A}}$ is qualitatively the same). Since the pulse remains azimuthally symmetric about the spine ($x=y=0$) throughout, the panel captures all of the transient features of the wave in the whole domain.

The pulse propagates at the local Alfv\'en speed $c_{A}=\sqrt{x^2+y^2+4z^2}$ along the magnetic fieldlines on which it is driven. The pulse is confined to the fieldlines throughout and does not propagate transverse to the field (as is characteristic of an Alfv\'en wave). The pulse therefore initially propagates down towards the null point, propagating primarily in the $z$-direction (i.e. moving nearly parallel to the spine, e.g., $t=0.75$, $t=1.25$). Subsequently, it begins to spread radially outward (from the spine) and ultimately tends towards propagating primarily radially outwards (i.e., mostly moving parallel to the fan-plane, e.g., $t=1.75$, $t=2.25$). 
From a 3D perspective, driving according to equation (\ref{pulse}) has introduced a toroidal pulse which undergoes isotropic dilation as it propagates travels down towards the fan. It propagates in the $\hat{z}$-direction at speed $c_{A}(0,0,z)=2z$ 
and dilates uniformly as magnetic fieldlines diverge from the spine. It accumulates near the fan-plane ($c_{A}(0,0,z)\rightarrow0$) where gradients in the pulse become increasingly steep and consequently, strong current builds. Eventually  resistivity becomes non-negligible, in agreement with previous wave-null studies (e.g., Galsgaard et al. \citeyear{klaus03}; McLaughlin \& Hood \citeyear{MH2004}), indicating that the fan-plane is a likely region for (wave-driven) heating events. 

We now consider fluid-velocity perturbations in the other orthogonal directions $\hat{\mathbf{B}}_{0}$ and $\hat{\mathbf{C}}$. All perturbations to variables in the $\hat{\mathbf{B}}_{0}$- and $\hat{\mathbf{C}}$-directions are of $\mathcal{O}(\alpha^2)$ and smaller, indicating that behaviour detailed in the following paragraphs is nonlinear and that the Alfv\'en wave is linearly decoupled from the magnetoacoustic waves (as first discussed in Thurgood \& McLaughlin \citeyear{Me2013_PMF}).

Figure \ref{vB_proper} shows the field-aligned perturbations in $|{v_B}|$, where we find a nonlinear $\mathcal{O}({\alpha^2}/2)$ disturbance that is everywhere cospatial to the Alfv\'en wave in ${v_{A}}$. The disturbance in $|{v_{B}}|$
 is not an independently propagating wave ($\beta=0$, hence such motion is prohibited $\mathbf{j}\times\mathbf{B}_{0}\cdot\mathbf{B}_{0}=0$). It is rather a consequence of the propagating Alfv\'en wave, which exerts a nonlinear magnetic pressure gradient along the magnetic field (\lq{ponderomotive force}\rq{}), which here is manifest in the longitudinal velocity perturbation.  This \emph{longitudinal daughter disturbance} has been identified as a general ponderomotive feature in nonlinear MHD and observed in nonlinear 2D wave-null studies (\citeyear{Me2013_NL2DAWAVE}; Thurgood \& McLaughlin \citeyear{Me2013_PMF}), and is equivalent to the \lq{ponderomotive wing\rq{} first reported in nonlinear 1.5D MHD by Verwichte et al. (\citeyear{Erwin99}).
{Via the ponderomotive force, Alfv\'en waves are (nonlinearly) compressible, which is manifest as these daughter disturbances. In sufficiently nonlinear regimes, this can result in enhanced dissipation of Alfv\'en wave energy (here, the nonlinearity is weak).} 
  We do not observe  the development of static perturbations to the longitudinal velocity. This indicates that nonlinear excitation of slow waves does not occur, again in agreement with the 2D results (in $\beta=0$, the slow mode has zero-speed and hence is manifest as a static, unstable perturbation, see Falle \& Hartquist \citeyear{FH02}).

In Figure \ref{vc_proper} we report the velocity component in the $\hat{\mathbf{C}}$-direction, which in $\beta=0$ is associated with fast magnetoacoustic waves and \emph{transverse daughter disturbances} (cross-field equivalents to the ponderomotive effect manifest in $|{v_{B}}|$). 
 During the driving phase, we observe the nonlinear generation of a multiple-lobed pulse of $\mathcal{O}({\alpha^2}/2)$, which subsequently propagates \emph{independently of the Alfv\'en wave}. The pulse travels towards the null point,  across fieldlines at speed $c_{A}$ (thus slowing on the approach to the null), crossing the fan and wrapping about the null point. 
Crucially, this wave crosses the fan plane (which is forbidden for the Alfv\'en wave). 
 This refraction along the Alfv\'en-speed profile and accumulation at the null point 
{ (where gradients and current eventually can become large leading to non-negligible  resistivity and heating)
}   has been documented as the typical feature of propagating fast magnetoacoustic waves at both 2D and 3D null points. After the driving phase, we find that there is no further excitation of fast magnetoacoustic waves.
 
The fast wave is excited nonlinearly by the process of ponderomotive mode excitation. The cross-field ponderomotive force exerted by an Alfv\'en wave is of the form
\begin{equation}
F_\perp = -\frac{1}{\mu \rho_0} \nabla_{\perp} \left(\frac{b_{z}^2}{2}\right)
\label{cfpmf}
\end{equation}
where  $\nabla_{\perp}$ is the cross-field spatial derivative, in this case equivalent to $\hat{\mathbf{C}}\cdot\nabla$, and $b_{z}=\pm\sqrt{\mu\rho_{0}}v_{z}$. Equation (\ref{cfpmf})  can be derived from equations  \ref{equation_MHD} using the process detailed in Thurgood \& McLaughlin (\citeyear{Me2013_PMF}) and is the non-dimensionalised equivalent to their equation (12).
Where transverse gradients in the Alfv\'en wave's amplitude are non-zero, the ponderomotive force acts across the field. When the average force over the wave's period is non-zero, fast magnetoacoustic waves are nonlinearly excited. Such non-zero average transverse gradients are typically assumed where the pulse propagates through regions of transverse inhomogeneity where $\nabla_{\perp} c_{A} \neq 0$ (phase-mixing regions).

%
 
Although the 3D proper null is transversely inhomogeneous throughout, we only observe the generation of fast waves during the driving phase and do not observe further excitation. This was previously reported for Alfv\'en waves at 2D nulls (Thurgood \& McLaughlin \citeyear{Me2013_NL2DAWAVE}). There are two possible explanations:
\begin{itemize}
\item The net ponderomotive force is non-zero, and acts to excite fast waves. However, after the initial driving period, a physical mechanism arises to suppress the further generation of fast waves. \\
\item The net ponderomotive force is zero throughout an no excitation of fast waves should occur. The excitation observed is an artefact due to driving at the boundary, which effectively  specifies a non-physical ponderomotive force during the driving phase. If this is the case, then the wave is a mathematical artefact as opposed to physical effect.\\
\end{itemize}
 As the 2D and 3D results agree, we refer the reader to section 4.2 of Thurgood \& McLaughlin (\citeyear{Me2013_NL2DAWAVE}) for a comprehensive discussion of possible explanations outlined above.

Finally,  we do not clearly observe any  cospatial disturbances manifest in $\hat{\mathbf{C}}$ (transverse daughter disturbances). Such a manifestation of the ponderomotive force is expected as the Alfv\'en wave assumes transverse gradients in the amplitude throughout. The transverse daughter cannot be distinguished as initially it is obscured by the fast wave, and at later times (when the Alfv\'en and fast waves are sufficiently separated) the configuration of the pulse is such that gradients across fieldlines rend the ponderomotive force exerted too small to observe at $\mathcal{O}(\alpha^2/2)$ scales.

\subsection{Improper Null, $\epsilon=0.5$}\label{section:3.2}
We now repeat the experiment and drive $v_A$ as per equation \ref{pulse} about an improper null point of eccentricity $\epsilon=0.5$. Here, the null is not azimuthally symmetric and fieldlines are predominantly aligned parallel to the $x$-axis. Again, we separately consider velocity in the $\hat{\mathbf{A}}$-, $\hat{\mathbf{B}}_0$-,and $\hat{\mathbf{C}}$-directions to isolate different modes of oscillation and disturbances.

Figure \ref{va_improper} shows the propagation of the Alfv\'en wave, manifest in $v_A$. As the azimuthal symmetry is lost for $\epsilon\neq1$, we show the figure in the planes defined by $x=0$ and $y=0$ (the right and left panels, respectively). 
As in $\S \ref{section:3.1}$ the pulse initially propagates down towards the null point, and is stretched  as fieldlines diverge from the spine line. Unlike in the proper-null case, the \lq{spreading}\rq{} effect is realised more rapidly in the $y=0$ plane than the $x=0$ plane. From a 3D perspective the dilation of the toroidal pulse is now anisiotropic, and the pulse which is initially a uniform, ring-shaped torus becomes prolate at later times, due to preferential stretching in the  $\hat{\mathbf{x}}$-direction (where fieldlines diverge at the greatest rate). The non-uniformity the dilation corresponds to the non-azimuthally symmetric manner in which field-lines diverge from running parallel to the spine line  when $\epsilon=0.5$ (see the right panel, Figure \ref{fig:fields}). 
Thus, like at the proper null, throughout the simulation (i.e., not just in these planes) the Alfv\'en wave propagates along fieldlines at the local Alfv\'en speed, and the difference between cases nulls of different epsilon $\epsilon$ is manifest in the differing rates of divergence in the $x=0$ and $y=0$ planes. 

Let us now consider the other directions $\hat{\mathbf{B}}_{0}$ and $\hat{\mathbf{C}}$. As per $\S \ref{section:3.1}$, we observe no linear disturbances to fluid-variables in these directions. Thus, the eccentricity of the null and departure from azimuthal symmetry does not facilitate any linear, geometric interaction between differing wave modes. 

 The fluid-velocity $|{v_{C}}|$ is shown in Figure \ref{vc_improper}. We see an independently propagating wave nonlinearly generated which propagates 
towards the null, crosses fieldlines and the fan, and wraps about the null point {where it eventually accumulates, as is typical of the fast wave}. The rate of refraction is again dependent on the Alfv\'en-speed profile, and therefore the  refraction effect is weaker in the $x=0$-plane (right panel) than the $y=0$-plane (left panel), corresponding to the relative steepness of the Alfv\'en-speed profile (see Paper 1). There is no further generation of independently propagating waves after the driving phase, as reported for the proper null case. 
 No daughter-type disturbances are clearly observed in $|{v_{C}}|$, as per $\S \ref{section:3.1}$.

The Alfv\'en wave was also found to be accompanied by a longitudinal daughter disturbance, manifest in 
in the $\hat{\mathbf{B}}_0$-direction
as per that detailed in \S \ref{section:3.1}, which is of $\mathcal{O}({\alpha^2}/2)$ and is everywhere cospatial to the Alfv\'en wave (hence we have not included an extra figure). As in the case of the proper null, no static disturbances (indicators of slow wave excitation) are observed in any of the components. 

We also considered intermediate values of $0\le\epsilon\le1$ and found qualitatively identical results. Wave behaviour about null points of different eccentricity only differs in that Alfv\'en waves spread throughout the region according to the differing rates of fieldline divergences, and any fast waves excited refract at differing rates according to the differing Alfv\'en-speed profiles. Crucially, different $\epsilon$ do not cause differing nonlinear effects. In all cases excitation of the fast wave occurs only during the driving phase and no subsequent conversion occurs.

\section{Conclusion}\label{section:4}
We have studied the behaviour of the Alfv\'en wave, and associated nonlinear effects, in the vicinity of both proper ($\epsilon=1$) and improper  ($\epsilon=0.5$) 3D magnetic null points. Independent of eccentricity, we find that:
\begin{itemize}
\item[$1.$] The Alfv\'en wave is confined to fieldlines, spreading radially outward as fieldlines diverge, and eventually assumes steep gradients as it nears the fan-plane,  causing current accumulation.
\\
\item[$2.$] The Alfv\'en wave continuously sustains a longitudinal daughter disturbance, due to its ponderomotive force directed along the equilibrium magnetic field $\hat{\mathbf{B}}_{0}$.
\\
\item[$3.$] During the driving phase,  independently propagating fast waves are introduced to the simulation. Subsequently, no further excitation of fast waves occurs.
\end{itemize}

The Alfv\'en wave behaves in its characteristic manner - it is confined to fieldlines and travels at the Alfv\'en speed, without exception. Here, where toroidal Alfv\'en wave pulses are driven about the spine, this results in an dilation/stretching of the toroid as the fieldlines diverge from the spine. When Alfv\'en waves are driven in forms such as that prescribed by equation (\ref{pulse}) at null points of differing eccentricity, the \emph{only} difference in the transient behaviour is the manner in which it expands,  which is entirely dependent on the topological eccentricity parameter $\epsilon$, which determines the degree of dilation anisotropy.

 This transient behaviour is wholly consistent with analytical predictions made by the 3D WKB approximation (McLaughlin et al. \citeyear{james083dwkb}), and the proper null results reported by Galsgaard et al. (\citeyear{klaus03}), and is an intuitive extension of the 2D results to 3D. As with the separatricies in the 2D models, in 3D the fan-surface is a location where current accumulates due to propagating Alfv\'en waves and is thus a  possible location of preferential, ohmic heating.

Nonlinearly, we also find that the theory of Alfv\'en waves at 2D magnetic null points (Thurgood \& McLaughlin \citeyear{Me2013_NL2DAWAVE}) carries over to the 3D case. Namely, in the manifestation of the daughter disturbances and the excitation of fast waves only during the driving phase. As in the 2D case, it is unclear whether the excitation of the fast wave is a physical effect or an artefact of the driving. The possibilities are fully discussed in Thurgood \& McLaughlin (\citeyear{Me2013_NL2DAWAVE}, section 4.2).

Overall, in the case of Alfv\'en waves at proper nulls, we find that changing $\epsilon$ only has minor, qualitative effects on the transient behaviour and that it does not effect the nature of the nonlinear effects. Despite the eccentricity and inhomogeneity, the wave retains its primary characteristic, that is, it is driven by magnetic tension only, and is a therefore a \emph{true} Alfv\'en wave as per Alfv\'en (\citeyear{Alfven42}).  The mere fact that a true Alfv\'en wave can exist about such a null point, indicates that the so-called \lq\lq{fully 3D}\rq\rq{} improper null point must contain some invariant direction, which is a necessary requirement for the existence of true Alfv\'en waves as discussed by Parker (\citeyear{Parker91}). Hence, null points of the form (\ref{eqn:3D_null_point}) must contain some form of invariance and thus the qualitative dynamics of the waves and their interaction is the same regardless of the degree of eccentricity. Choices of $\epsilon$ give the following forms of invariance:
\begin{itemize}
\item $\epsilon=0$ yields an Cartesian/translational invariance (of the equilibrium field) $\partial B/\partial z=0$ .\\
\item $\epsilon=1$ yields an azimuthal/torsional invariance $\partial B / \partial \theta =0$.\\
\item $0< \epsilon \le 1$ yields a general torsional invariance $\partial B / \partial S = 0$ where S is a direction around the spine which is associated with the flux-function.
\end{itemize}

The previously-studied 2D null point models correspond to the special case of $\epsilon=0$, where the torsional symmetry tends to translational symmetry. As invariance is intimately connected to the concept of the Alfv\'en wave, the extent of the similarity and difference between translational and rotational invariances determines the extent to which the theory of Alfv\'en waves at 2D nulls carries across to 3D nulls. Rotational symmetry ($\epsilon>0$), imposes restrictions on the geometric form that Alfv\'en wave pulses may take, unlike in 2D modelling (or, equivalently, the translationally symmetric 3D case $\epsilon=0$).  As an Alfv\'en wave must perturb fluid variables across the field without causing a (linear) magnetic pressure gradient to arise, in cases of rotational invariance Alfv\'en waves \emph{are always torsional} (one could consider twisting along isosurfaces of $\hat{\mathbf{A}}$). Thus, any effects noted in 2D wave-null modelling which are geometry-specific  (e.g., many 2D models consider planar Alfv\'en waves only) will not necessarily extend to 3D in as straightforward a manner as we have found for the cases considered in this paper.

\section{Summary}\label{section:4.1}

The primary aim of the study presented here and in  Paper 1  was to address the question of how the theory of MHD waves at 2D null points extends to fully 3D {potential} null points. Summarising, these studies have shown that the transient behaviour of the fast and Alfv\'en waves are as follows:
\begin{itemize}
\item{\textbf{Fast Waves:} Propagate according to the Alfv\'en-speed profile along and across the fieldlines, refracting and accumulating at the null point. Different $\epsilon$ simply alters the rate of refraction (due to steeper/gentler Alfv\'en-speed profiles). Independent of eccentricity, the wave eventually accumulates at the null point itself. }
\\
\item{\textbf{Alfv\'en Waves:} Are confined to fieldlines, travelling at $c_{A}$, and thus pulses exhibit a \lq{spreading}\rq{} effect as fieldlines diverge, typically accumulating near the fan plane. Different parameters $\epsilon$ simply change the preferential direction on the pulse spreading, due to the manner in which fieldlines diverge from the spine line.}
\end{itemize}

Thus, in terms of transient behaviour the 2D result that different modes accumulate in predictable parts of null point topology extends completely. Additionally, we report that the  nonlinear effects of the Alfv\'en and fast waves are consistent with the 2D models; namely, the sustaining of daughter disturbances and the excitation of fast waves due to the ponderomotive force of an Alfv\'en wave.
Overall, for the $\beta=0$, {potential null} case, \emph{the 2D theory carries over in an intuitive way.}


{This paper has considered Alfv\'en wave behaviour about proper and improper potential null points. However, we finish by noting that solar nulls are unlikely to be current-free (potential), as is  found to be the case in field extrapolations (e.g., Valori et al. \citeyear{2012SoPh..278...73V}, Sun et al. \citeyear{2012ApJ...757..149S}). 
Non-potential nulls are more topologically complicated and asymmetric than the potential class (examples of non-potential field line geometry can be found in, e.g., Al Hachani et al. \citeyear{2010A&A...512A..84A}, Pontin et al. \citeyear{2011A&A...533A..78P}, and Pontin \citeyear{2011AdSpR..47.1508P}).
 Currently, it is unclear to what extent the above results hold in the case of non-potential null points. It may well be expected that the fast wave will refract about the Alfv\'en-speed profile (regardless of field line structure) as in Paper 1, however the increased asymmetry could possibly complicate the Alfv\'en wave dynamics at such nulls. Indeed, given the discussion of invariance being a requirement for the existence of a  true Alfv\'en wave, does the asymmetry prohibit this mode entirely? Additionally, the greater inhomogeneity may increase the efficiency of any ponderomotive mode excitation. Wave behaviour may be complicated further still if such nulls are embedded in large scale quasi-separatrix layers, as in Masson et al. (\citeyear{2009ApJ...700..559M}). As nulls in the solar atmosphere will be non-potential, the further extension of 3D wave-null theory from  potential to non-potential cases is an important outstanding question and should be addressed in future research. 
}

\begin{acknowledgements}
The authors acknowledge IDL support provided by STFC. JOT acknowledges travel support provided by the RAS and the IMA, and a Ph.D. scholarship provided by Northumbria University. The computational work for this paper was carried out on the joint STFC and SFC (SRIF) funded cluster at the University of St Andrews (Scotland, UK).
\end{acknowledgements}

\bibliographystyle{aa} 
\bibliography{references.bib}

\end{document}